\documentclass[aps,prb,floatfix,twocolumn]{revtex4}
\usepackage{graphics}
\usepackage{dcolumn}
\usepackage{epsfig}

\newcommand{\bc}{{\bf c }}

\newcommand{\bk}{{\bf k }}

\newcommand{\bp}{{\bf p }}

\newcommand{\br}{{\bf r }}

\newcommand{\bB}{{\bf B}}

\newcommand{\bK}{{\bf K}}

\begin{document}

\pagenumbering{arabic}

\title
{Making Massless Dirac Fermions from a Patterned\\ Two-Dimensional Electron Gas}
\author{Cheol-Hwan Park}
\author{Steven G. Louie}
\email{sglouie@berkeley.edu}
\affiliation{Department of Physics, University of California at Berkeley,
Berkeley, California 94720\\
Materials Sciences Division, Lawrence Berkeley National Laboratory,
Berkeley, California 94720}

\date{\today}

\begin{abstract}
Analysis of the electronic structure of an ordinary
two-dimensional electron gas (2DEG)
under an appropriate external periodic potential of
hexagonal symmetry reveals that massless Dirac fermions
are generated near the corners of the supercell Brillouin zone.
The required potential parameters are found to be achievable
under or close to laboratory conditions.
Moreover, the group velocity is tunable by changing either
the effective mass of the 2DEG or the lattice parameter
of the external potential, and it
is insensitive to the potential amplitude.
The finding should provide a new class of systems
other than graphene for
investigating and exploiting massless Dirac fermions
using 2DEGs in semiconductors.
\end{abstract}

\maketitle

Graphene~\cite{novoselov:2005PNAS_2D,novoselov:2005Nat_Graphene_QHE,
zhang:2005Nat_Graphene_QHE,berger:2006Graphene_epitaxial},
a honeycomb lattice of carbon atoms, is composed
of two equivalent sublattices of atoms.
The dynamics of the low-energy charge carriers in graphene may be
described to a high degree of accuracy by a massless Dirac equation
with a two-component pseudospin
basis which denotes the amplitudes of the electronic states
on these two sublattices.
The quasiparticles
have a linear energy dispersion near the corners $\bK$ and $\bK'$
(the Dirac points) of the hexagonal Brillouin
zone~\cite{wallace:1947PR_BandGraphite,PhysRev.104.666,ando:1998JPSJ_NT_Backscattering,ando:1998JPSJ_NT_BerryPhase}.
Consequently, the density of states (DOS) varies linearly and
vanishes at the Dirac point energy.
The sublattice degree of freedom of the wavefunctions is given by
a pseudospin vector that is either parallel or anti-parallel
to the wavevector measured from the Dirac point, giving rise to
a chirality being 1 or $-1$, respectively~\cite{wallace:1947PR_BandGraphite,ando:1998JPSJ_NT_Backscattering,ando:1998JPSJ_NT_BerryPhase}.
These two fundamental properties of graphene, linear energy dispersion and the chiral nature of the
quasiparticles, result in interesting phenomenon such as half-integer quantum Hall
effect~\cite{novoselov:2005Nat_Graphene_QHE,zhang:2005Nat_Graphene_QHE},
Klein paradox~\cite{katsnelson:2006NatPhys_Graphene_Klein},
and suppression of backscattering~\cite{ando:1998JPSJ_NT_Backscattering,
ando:1998JPSJ_NT_BerryPhase,mceuen:1999PRL_NT_Backscattering},
as well as some novel predicted properties such as electron
supercollimation in graphene
superlattices~\cite{park:2008NatPhys_GSL,park:2008NL_Supercollimation,park:126804}.

As a possible realization of another two-dimensional (2D) massless Dirac particle system,
theoretical studies on the physical properties of particles
in optical honeycomb lattices~\cite{PhysRevLett.70.2249} have been
performed~\cite{zhu:260402,wu:070401,wu:arxiv2007,shao:prl,haddad:arxiv2008}.
The behaviors of ultra-cold atoms in a honeycomb lattice potential
were considered, in principle, to
be equivalent to those of the low-energy charge carriers in graphene~\cite{zhu:260402}.

In this Letter, we propose a different practical
scheme for generating massless Dirac fermions.
We show through exact numerical calculations within an independent particle picture
that applying an appropriate nanometer-scale periodic potential with hexagonal
symmetry onto conventional two-dimensional electron gases (2DEGs) will
generate massless Dirac fermions at the corners of the supercell Brillouin zone (SBZ).
We find that the potential configurations needed should be within or close to
current laboratory capabilities, and this approach could benefit from
the highly developed experimental techniques of 2DEG physics~\cite{book:2DEG}
including recent advances in self-assembly
nanostructures~\cite{PhysRevB.58.1506,PhysRevLett.82.996,ribeiro:microel}.

We moreover find that the band velocity and the energy window within which the dispersion is linear
may be varied by changing the superlattice parameters or the effective mass
of the host 2DEG, thus providing a different class of massless Dirac fermion systems
for study and application.
Interestingly, the amplitude of the periodic potential does not
affect the band velocity about the Dirac points.
But, if the external periodic potential is too weak,
there is no energy window within which the DOS vanishes linearly.
The linear energy dispersion and the chiral nature of states
around the Dirac points of these 2DEG superlattices are
found to be identical to those of graphene.
Also, the associated up and down pseudospin states naturally correspond to
states localizing, respectively, on two equivalent sublattice sites
formed by the superlattice potential.

In order to investigate the properties of charge carriers in 2DEGs
under an external periodic potential,
we work within the one-electron framework
with realistic material parameters. However, the point of this Letter
is not the methodology itself, but
the new idea and its feasibility.
The study makes a connection between two of the largest and most active fields in condensed
matter physics and nanoscience these days, graphene and 2DEGs,
and provides experimentalists working on 2DEGs
with a previously unexploited way of generating a new class of 2D
massless Dirac
fermions and of tuning their properties.

Let us consider a 2DEG with
$E(\bp)=p^2/2m^*$ where $m^*$ is the
electron effective mass.
This is a good approximation to the energy dispersion of the
lowest conduction band in diamond- or zinc-blende-type semiconductor
quantum wells, which have
effective masses ranging from
$m^*=0.02~m_e\sim0.17~m_e$ where $m_e$ is the free electron
mass~\cite{yu:book}.
We shall consider explicitly two cases in the numerical calculations:
$m^*=0.02~m_e$ and $m^*=~0.04~m_e$.

  \begin{figure}
  \includegraphics[width=1.0\columnwidth]{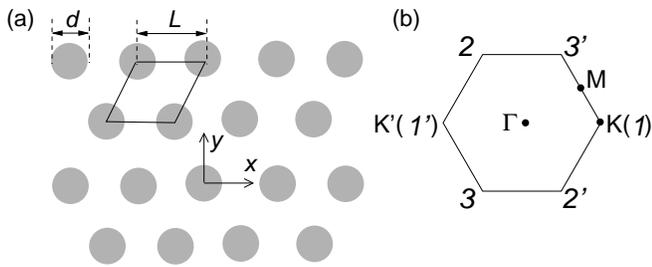}
  \caption{(a) A muffin-tin type of hexagonal
  periodic potential with a spatial period $L$. The potential is
  $U_0\,(>0)$ inside the gray disks with diameter $d$ and zero outside.
  (b) The Brillouin zone of hexagonal lattice in (a).}
  \label{Fig1}
  \end{figure}

Figure~\ref{Fig1}(a) shows the muffin-tin periodic potential considered
in our numerical calculations,
whose value is $U_0\,(>0)$ in a triangular array of disks of diameter $d$
and zero outside. Figure~\ref{Fig1}(b) is the corresponding
Brillouin zone. The muffin-tin form is chosen for ease of discussion;
the conclusions presented here are generally valid for any hexagonal
potential.

  \begin{figure}
  \includegraphics[width=1.0\columnwidth]{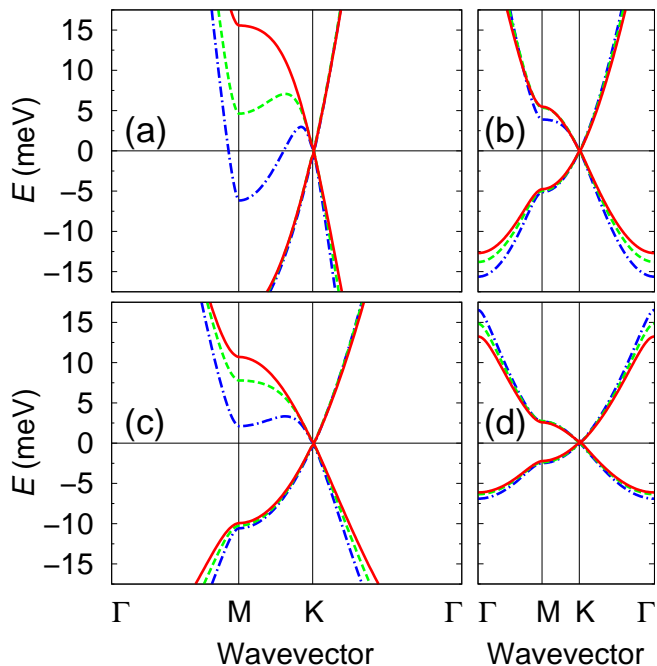}
  \caption{(color online). Calculated energy bandstructure of the lowest two bands of
  a hexagonal 2DEG superlattice shown in Fig.~\ref{Fig1}(a).
  (a) $m^*=0.02\,m_e$ and $L=20$~nm,
  (b) $m^*=0.02\,m_e$ and $L=40$~nm,
  (c) $m^*=0.04\,m_e$ and $L=20$~nm, and
  (d) $m^*=0.04\,m_e$ and $L=40$~nm.
  The diameter of the disks $d$ is set to $d=0.663\,L$~(see text).
  Solid red, dashed green, and dash-dotted blue lines show results for $U_0$ equal to
  200~meV, 100~meV, and 50~meV, respectively. The Dirac point energy
  (i.\,e.\,, the energy at the crossing of the two bands at $\bK$) is set to zero.}
  \label{Fig2}
  \end{figure}

We shall first discuss our numerical results and later consider
the approximate analytic solutions.
Figure~\ref{Fig2} shows the calculated bandstructures of the lowest two bands for
several hexagonal 2DEG superlattices with different effective
mass $m^*$, lattice parameter $L$ (with potential barrier diameter
$d=0.663\,L$~\cite{note:W}), and barrier height $U_0$.
As the barrier height is decreased, the energy window within which
the energy dispersion is linear is reduced (Fig.~\ref{Fig2}).
(The potential barrier heights used in our calculations are typical
of values employed in confining 2DEGs~\cite{book:2DEG}.)
However, the group velocity at the Dirac point is
insensitive to $U_0$ (Fig.~\ref{Fig2}).
But, as $m^*$ or {\it L} is increased,
the group velocity decreases (Fig.~\ref{Fig2}).

  \begin{figure}
  \includegraphics[width=0.7\columnwidth]{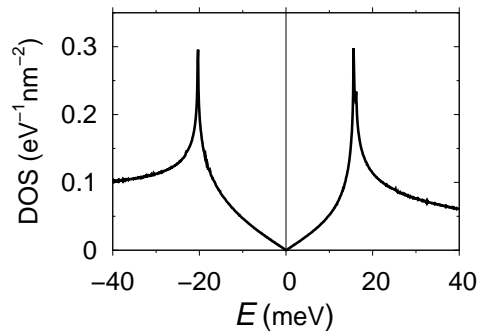}
  \caption{The DOS of a triangular 2DEG superlattice
  with $m^*=0.02\,m_e$, $U_0=200$~meV, $L=20$~nm and $d=13.3$~nm.
  (The zero of energy is set at the Dirac point.)
  The charge density needed to fill the
  conduction band up to the Dirac point energy is
  $5.7\times10^{11}~{\rm cm}^{-2}$.}
  \label{Fig3}
  \end{figure}

Figure~\ref{Fig3} shows the DOS of a hexagonal 2DEG superlattice
with $m^*=0.02\,m_e$, $U_0=200$~meV, $L=20$~nm and $d=13.3$~nm.
The DOS has a linear behavior around the Dirac point energy
within a $\sim30$~meV energy window. The charge density
required to dope the system to reach the Dirac point energy is $5.7\times10^{11}~{\rm cm}^{-2}$,
which is in the range of typical value in 2DEG studies~\cite{book:2DEG},
and may be tuned by applying a gate voltage or by light illumination~\cite{PhysRevLett.83.2234}.

With the above results established from the numerical calculations,
to gain further insight, we now present
analytical expressions for the energy dispersion relation and
wavefunctions around the SBZ corners obtained from degenerate perturbation theory.
We concentrate on states with wavevector $\bk+\bK$
near the $\bK$ point in the SBZ, i.\,e.\,, $|\bk|\ll |\bK|$.
Let us set the energy of the empty lattice bandstructure at the
{\bf K} point to zero, define
$W$ as the Fourier component of the periodic potential connecting
${\it 1}\to{\it 2}$, ${\it 2}\to{\it 3}$ and ${\it 3}\to{\it 1}$
in Fig.~\ref{Fig1}(b)~\cite{note:W},
and denote $v_0$ as the group velocity of the electron state at the $\bK$ point
of the 2DEG before applying the periodic potential.
[For $E(\bp)=p^2/{2m^*}$, $v_0=\hbar K/m^*$.
However, the derivation below is not confined to a quadratic energy dispersion
for the original 2DEG.]
Due to the inversion symmetry of the system considered here,
$W$ is real.
The wavefunction $\psi_\bk(\br)$ may be approximately expressed as a linear combination
of three planewave states
\begin{eqnarray}
&&\psi_\bk(\br)=\frac{1}{\sqrt{3A_c}}\left[c_{\it 1}\exp\left(i(\bK_{\it 1}+\bk)\cdot\br\right)\right.
\nonumber \\
&&\left.+c_{\it 2}\exp\left(i(\bK_{\it 2}+\bk)\cdot\br\right)+c_{\it 3}\exp\left(i(\bK_{\it 3}+\bk)\cdot\br\right)\right]\,,\nonumber\\
\label{eq:psi}
\end{eqnarray}
where $A_c$ is the area of the 2DEG and $\bK_{\it 1}$, $\bK_{\it 2}$ and $\bK_{\it 3}$ represent
wavevectors at the
SBZ corners {\it 1}, {\it 2} and {\it 3}, respectively,
in Fig.~\ref{Fig1}(b).
Equivalently, we could express the eigenstate
as a three-component column vector
$\bc=(c_{\it 1}, c_{\it 2}, c_{\it 3})^{\rm T}$.
Within this basis,
the Hamiltonian $H$, up to first order in $k$, is given by
$H=H_0+H_1$,
where
\begin{equation}
H_0=W\left( \begin{array}{ccc}
0 & 1 & 1 \\
1 & 0 & 1 \\
1 & 1 & 0
\end{array} \right)
\label{eq:H0}
\end{equation}
and
\begin{equation}
H_1=\hbar v_0\,k\left( \begin{array}{ccc}
\cos\theta_\bk & 0 & 0 \\
0 & \cos\left(\theta_\bk-\frac{2\pi}{3}\right) & 0 \\
0 & 0 & \cos\left(\theta_\bk-\frac{4\pi}{3}\right) \end{array} \right)\ .
\label{eq:H1}
\end{equation}
Here $\theta_\bk$ is the polar angle of the wavevector $\bk$ from the $+x$ direction.

The eigenvalues of the unperturbed Hamiltonian  $H_0$ are
\begin{equation}
E_0=-W,\ -W,\ 2W\,,
\label{eq:E0}
\end{equation}
which are also the energies of the states of the superlattice at $\bk=0$.
We now focus on the doubly-degenerate eigenstates with eigenvalue $-W$.
We shall find the {\bf k}-dependence of the
eigenenergies and eigenvectors of $H$ corresponding to these two states
within degenerate perturbation theory by treating $H_1$ as a perturbation,
which is approximate for $\hbar\,v_0\,k<W$.
Also, for a wavefunction in the form of Eq.~(\ref{eq:psi})
to give a good description of the
actual wavefunction, the energy to the next planewave state
[which is $\hbar^2(2K)^2/2m^*-\hbar^2K^2/2m^*$]
should be smaller than $W$.
Therefore, the approximation is valid within
the regime
$\hbar\,v_0\,k< W<\frac{3\hbar^2K^2}{2m^*}$,
or, equivalently,
\begin{equation}
\frac{4\pi\hbar^2}{3m^*L}k< W<\frac{8\pi^2\hbar^2}{3m^*L^2}\,,
\label{eq:criterion}
\end{equation}
where we have used $K=4\pi/3L$.

The two eigenvectors of $H_0$ with eigenvalue $-W$ are
\begin{equation}
\bc_1=\frac{1}{\sqrt{2}}\left(
\begin{array}{r}
0\\
1\\
-1
\end{array}
\right)\ \ \text{and}\ \ \
\bc_2=\frac{1}{\sqrt{6}}\left(
\begin{array}{r}
2\\
-1\\
-1
\end{array}
\right)\,.
\label{eq:v1v2}
\end{equation}
The term $H_1$, when restricted to the sub-Hilbert-space spanned by
the two vectors in Eq.~(\ref{eq:v1v2}), is represented
by a $2\times2$ matrix $\tilde{H_1}$
\begin{equation}
\tilde{H_1}=\hbar\frac{v_0}{2}
\left(
\begin{array}{rr}
-k_x & -k_y\\
-k_y & k_x
\end{array}
\right)\,,
\label{eq:H1tilde}
\end{equation}
where $k_x=k\cos\theta_\bk$ and $k_y=k\sin\theta_\bk$.
After a similarity transform $M=U^\dagger \tilde{H_1} U$ with
\begin{equation}
U=\frac{1}{2}
\left(
\begin{array}{rr}
1+i\ \  & -1-i\\
-1+i\ \  & -1+i
\end{array}
\right)\,,
\label{eq:U}
\end{equation}
we obtain
\begin{equation}
M=\hbar\frac{v_0}{2}
\left(
k_x\sigma_x+k_y\sigma_y
\right)\,.
\label{eq:M}
\end{equation}
Here, $\sigma_x$ and $\sigma_y$ are the Pauli matrices.
Equation~(\ref{eq:M}) is just the effective Hamiltonian of
graphene~\cite{wallace:1947PR_BandGraphite} with a group velocity
\begin{equation}
v_g=v_0/2=\frac{\hbar K}{2m^*}=\frac{2\pi\hbar}{3m^*L}\,.
\label{eq:vg}
\end{equation}
Therefore, the group velocity is reduced if the effective mass is increased
or the lattice parameter of the superlattice is increased, but it is
insensitive to the amplitude of the external periodic potential,
which explains the results of the numerical calculations shown in Fig.~\ref{Fig2}.
Also, the size of the linear energy dispersion window [Eq.~(\ref{eq:criterion})]
is dictated by the value of {\it W}, which for the muffin-tin
potential in Fig.~\ref{Fig1}(a) with $d=0.663\,L$ is
$W=0.172\,U_0$~\cite{note:W} in agreement with the numerical calculations~\cite{note:imperfection}.

The eigenvalues and the eigenvectors of $M$ are
\begin{equation}
E(s,\bk)=s\,\hbar \frac{v_0}{2}\,k\,,
\label{eq:E_k}
\end{equation}
and
\begin{equation}
\left|s,\bk\right>=\frac{1}{\sqrt{2}}
\left(\begin{array}{c}
1\\0\end{array}\right)+
\frac{1}{\sqrt{2}}\,s\,e^{i\theta_\bk}
\left(\begin{array}{c}
0\\1\end{array}\right)\,,
\label{eq:M_psi}
\end{equation}
respectively, where $s=\pm1$ is a band index~\cite{wallace:1947PR_BandGraphite}.
The vectors $\left(\begin{array}{c}1\\0\end{array}\right)$
and $\left(\begin{array}{c}0\\1\end{array}\right)$
are the up and the down pseudospin eigenstates of $\sigma_z$,
respectively~\cite{note:Kp}.

  \begin{figure}
  \includegraphics[width=1.0\columnwidth]{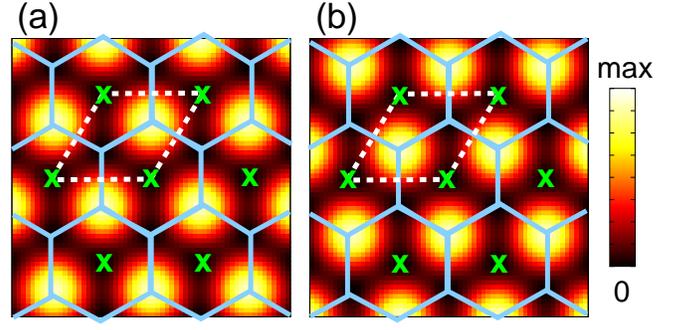}
  \caption{(color online). Probability densities of the pseudospin states
  in a hexagonal 2DEG superlattice
  (a) $\left|\left<\br|\uparrow\right>\right|^2$ and
  (b) $\left|\left<\br|\downarrow\right>\right|^2$.
  Note that the amplitudes of the states
  are localized at two different but equivalent sublattices. The centers of the
  potential barrier disks [Fig.~\ref{Fig1}(a)] are shown as `x' marks,
  and the honeycomb structure is drawn to illustrate the connection to the
  superlattice structure.}
  \label{Fig4}
  \end{figure}

The up and the down pseudospin eigenstates may be expressed
within the basis of the original Hamiltonian $H$ using Eqs.~(\ref{eq:U}) and~(\ref{eq:v1v2}) as
\begin{eqnarray}
\left|\uparrow\right>
&=&\frac{1}{\sqrt{3}}e^{i\frac{3\pi}{4}}\left(1,\ e^{-i\frac{2\pi}{3}},\ e^{-i\frac{4\pi}{3}}
\right)^{\rm T}
\label{eq:up}
\end{eqnarray}
and
\begin{eqnarray}
\left|\downarrow\right>
&=&\frac{1}{\sqrt{3}}e^{i\frac{3\pi}{4}}\left(1,\ e^{i\frac{2\pi}{3}},\ e^{i\frac{4\pi}{3}}
\right)^{\rm T}\,,
\label{eq:down}
\end{eqnarray}
respectively~\cite{note:c}.
The real space pseudospin wavefunctions
$\left<\br|\uparrow\right>$ and $\left<\br|\downarrow\right>$
are obtained by
putting the coefficients in Eqs.~(\ref{eq:up}) and~(\ref{eq:down})
into Eq.~(\ref{eq:psi}).
Figures~\ref{Fig4}(a) and~\ref{Fig4}(b) show $\left|\left<\br|\uparrow\right>\right|^2$
and $\left|\left<\br|\downarrow\right>\right|^2$, respectively.
Note that the up and the down pseudospin states are seen as localized at
one of the two
equivalent sublattices formed by the external periodic potential,
in perfect analogy with the behavior in graphene.

Let us now consider the Landau levels for the above hexagonal 2DEG superlattice
in a magnetic field $\bB=B\,\hat{z}$ when the Fermi level is at the Dirac point
energy.
In exact analogy with graphene, the low-energy Landau levels
shifted by the energy $W$
(i.\,e.\,, having the energy zero at the Dirac point energy)
are $E_n=\hbar\omega_c\,{\rm sgn}(n)\sqrt{|n|}$,
where $\omega_c=v_0\sqrt{|e|B/2\hbar c}$
is the cyclotron frequency~\cite{PhysRev.104.666,note:LL}.
Here, for an appropriately constructed superlattice potential,
the half-integer quantum Hall effect~\cite{novoselov:2005Nat_Graphene_QHE,
zhang:2005Nat_Graphene_QHE} should be observable.

In conclusion, we have shown that chiral massless Dirac fermions
are generated if an appropriate nanometer-scale periodic
potential with hexagonal symmetry is applied to a conventional 2DEG
in semiconductors. These quasiparticles have a linear
energy dispersion, with a group velocity half the value
of the states before applying the periodic potential, and
a wavefunction whose chiral structure exactly the same as that of graphene.
The up and the down pseudospin states are shown
to be localized at two different but equivalent sublattices
formed by the superlattice potential.
The quasiparticle group velocity moreover is tunable by changing the
effective mass of the original 2DEG
or the lattice parameter of the superlattice potential.
Our findings thus provide a new class of systems for experimental investigations
and practical applications of 2D massless Dirac quasiparticles.

We thank
Kathryn Todd, Ileana Rau, Sami Amasha,
Philip Kim, Yunchul Chung, Jiwoong Park,
and Jannik Meyer for fruitful discussions.
This work was supported by NSF Grant
No. DMR07-05941 and by the Director, Office of Science, Office of Basic Energy
Sciences, Division of Materials Sciences and Engineering Division,
U.S. Department of Energy under Contract No. DE- AC02-05CH11231.
Computational resources have been provided by NPACI and NERSC.

{\it Note added in proof.} -- After we finished writing up
this manuscript, we became aware of Ref.~\cite{wunsch:njp} in which
a similar methodology to the one developed independently here
was developed. The objects of investigation
(which are cold fermionic atoms confined in honeycomb optical lattices)
and the main ideas and conclusions in Ref.~\cite{wunsch:njp},
however, are qualitatively different from those in this Letter.


\begin{thebibliography}{34}
\bibitem[{not({\natexlab{a}})}]{note:W}
\bibinfo{note}{The Fourier component $W$, defined to be the value of the
  external periodic potential [Fig.~\ref{Fig1}(a)] in Fourier space at the
  smallest reciprocal lattice vector, is given by
  $W=\frac{d}{2L}J_1\left(\frac{2\pi d}{\sqrt{3}L}\right)U_0$ where $J_1(x)$ is
  the first-order Bessel function. For a given $U_0$, the maximum value of $W$
  is $W=0.172\,U_0$ at $d=0.663\,L$. In our numerical calculations, we take
  advantage of this. For comparison, if $d=0.5\,L$, $W=0.145\,U_0$}.
\bibitem[{not({\natexlab{c}})}]{note:Kp}
\bibinfo{note}{Similar procedures can be applied to find the energy and the
  wavefunction of massless Dirac fermions with wavevectors near $\bK'$
  [Fig.~\ref{Fig1}(b)]}.
\bibitem[{not({\natexlab{d}})}]{note:c}
\bibinfo{note}{Putting Eqs.~(\ref{eq:up}) and~(\ref{eq:down}) into
  Eq.~(\ref{eq:M_psi}), we obtain, within a phase factor, the eigenvector in
  the original planewave coefficient basis [Eq.~(\ref{eq:psi})] $$
  \bc(s,\bk)=\left(\begin{array}{l} \sqrt{\frac{2}{3}} \left( \begin{array}{l}
  \vspace{0.05in} \cos\left(\frac{\theta_\bk}{2}\right)\\ \vspace{0.05in}
  \cos\left(\frac{\theta_\bk}{2}+\frac{2\pi}{3}\right)\\
  \cos\left(\frac{\theta_\bk}{2}+\frac{4\pi}{3}\right) \end{array} \right)\ \
  (s=1)\\ \\ \\ \sqrt{\frac{2}{3}} \left( \begin{array}{l} \vspace{0.05in}
  \sin\left(\frac{\theta_\bk}{2}\right)\\ \vspace{0.05in}
  \sin\left(\frac{\theta_\bk}{2}+\frac{2\pi}{3}\right)\\
  \sin\left(\frac{\theta_\bk}{2}+\frac{4\pi}{3}\right) \end{array} \right)\ \
  (s=-1) \end{array}\,\right. $$}.
\bibitem[{not({\natexlab{e}})}]{note:LL}
\bibinfo{note}{The expression for the (relativistic) Landau levels is valid
  when the spacing between Landau levels are much smaller than the size of the
  energy window within which the dispersion is linear, regardless of the
  existence of higher energy states, just as in graphene. For example, in a
  triangular 2DEG superlattice with $m^*=0.02\,m_e$, $U_0=200$~meV, $L=20$~nm
  and $d=13.3$~nm, this condition is equivalent to $B<2$~T}.
\end{thebibliography}
\end{document}